\documentclass[a4paper]{article}

\usepackage{INTERSPEECH2018}

\usepackage{tipa}
\usepackage{multirow}

\title{Subjective and objective experiments on the influence of speaker's gender on the unvoiced segments
}

\name{A Madhavaraj$^1$, T V Ananthapadmanabha$^2$ and A G Ramakrishnan$^1$}
\address{
  $^1$MILE Lab, Department of Electrical Engineering, Indian Institute of Science, Bangalore 560012\\  $^2$Voice and Speech Systems, Malleswaram, Bangalore 560003, India
}
\email{madhavaraja@iisc.ac.in, tva.blr@gmail.com, agr@iisc.ac.in}

\begin{document}

\maketitle	

\begin{abstract}

Subjective and objective experiments are conducted to understand the extent to which a speaker's gender influences the acoustics of unvoiced (U) sounds. U segments of utterances are replaced by the corresponding segments of a speaker of opposite gender to prepare modified utterances. Humans are asked to judge if the modified utterance is spoken by one or two speakers. The experiments show that human subjects are unable to distinguish the modified from the original. Thus, listeners are able to identify the U segments irrespective of the gender, which may be based on some speaker-independent invariant acoustic cues. To test if this finding is purely a perceptual phenomenon, objective experiments are also conducted. Gender specific HMM based phoneme recognition systems are trained using the TIMIT training set and tested on (a) utterances spoken by the same gender (b) utterances spoken by the opposite gender and (c) the modified utterances of the test set. As expected, the performance is the highest for case (a) and the lowest for case (b). The performance degrades only slightly for case (c). This result shows that the speaker's gender does not as strongly influence the acoustics of U sounds as they do the voiced sounds.

\end{abstract}	

\noindent\textbf{Index Terms}: Unvoiced sounds, voiced sounds, perception, vocal tract, phoneme recognition, gender-dependence, stops.

\section{Introduction}
It has been shown that the formant data of vowel sounds are strongly influenced by the speaker's gender and age \cite{pb,fant}. From a speech production point of view, a change in size of the vocal tract must affect the vocal tract frequency response of unvoiced sounds as well. Such a view is supported by several previous studies. The influence of gender on the perception of unvoiced sounds has been reported \cite{sh-s}. Using linear prediction coefficients (LPCs) based features, the highest accuracy for gender recognition has been obtained for voiced fricatives of all the speech sounds \cite{gender}. Gradient normalization has been applied on fricative sounds in a gender identification perception experiment \cite{fric}. To compensate for the speaker's influence, spectral warping techniques \cite{LandR96, UandW99} have been proposed and techniques developed for voiced (V) sounds are applied to the unvoiced (U) sounds as well. Speaker recognition accuracy is known to improve if U sounds are handled separately, indicating speaker specific influence on U sounds \cite{spk_uv}.

In \cite{reldiscphones}, the relative speaker discriminative properties of different phones are studied and the authors have found that the stop consonants provide the worst performance. Huang and Epps \cite{huang_forensics} have integrated phone segmentation with forensic voice comparison (FVC) system and showed that vowels and nasals contribute the most to the reliability and validity of the system. Recently, Moez et. al. \cite{moez_forensics} have investigated the impact of each phoneme class for FVC application. They showed that vowels and nasals perform better than averaged phonemic content, while the fricatives do not perform better than the averaged content. The speaker recognition experiments of Jung et. al. \cite{jungspk} indicate that the affricate, stop and fricative phone classes do not contribute much in recognition and by selecting lesser number of frames from these classes improves the recognition performance.

Since there is no clearly defined formant structure for unvoiced sounds, we attempt to deduce the influence of gender on U sounds indirectly by means of subjective and objective experiments. The proposed experiments also indirectly look at the relative importance of modeling the human speech production and perception mechanisms in designing speech and speaker recognition systems.

This paper is organized as follows. Section 2 describes the algorithm used for generating the stimuli by mixing unvoiced and voiced segments from the recordings of the same sentence spoken by two speakers from the opppsite sexes. It also discusses the listening test conducted on 50 human subjects for subjective evaluation of our hypothesis. In Sec. 3, we illustrate the ASR experiments that have been conducted by training independent phoneme recognition models on male and female recordings and testing those models on the different types of modified utterances produced. Section 4 describes the results of the subjective and objective experiments conducted in our studies. Finally, we conclude the paper in Sec. 5.

\section{Subjective Listening Experiments}

\subsection{Preparation of the modified utterances as stimuli}
The TIMIT \cite{timit} database gives the boundaries of voiced sonorants and unvoiced segments (silence, fricatives and stops). All vowels, semi-vowels and nasals are included under the voiced sonorant phones. Short SX utterances of the TIMIT database, with 4 to 6 words, are used for the perceptual tests. Each original utterance is modified in two ways: (a) Set1 of test utterances: All unvoiced segments in the utterance (of an SX file) spoken by a male speaker are replaced by the corresponding (i.e., from the same SX filename) unvoiced segments of a female speaker (denoted as M$<$FU) and vice-versa (F$<$MU). For this set, five original utterances from each gender are used. (b) Set2 of test utterances: In addition to the replacement of all the unvoiced segments, as in Set1, two consecutive sonorant segments of the utterance spoken by a male speaker are replaced by the corresponding sonorant segments of a female speaker (denoted as M$<$FvU) and vice-versa (F$<$MvU). For this set, another five utterances are used. The Set2 stimuli utterances are included partly to increase the level of confidence of the listeners (that there are indeed utterances with mixed speakers) and partly to study the relative perceptual influence of speaker's identity on the voiced and unvoiced segments.

The procedure to mix two signals is not a straight-forward task, since the same sentence spoken by two speakers may have different phonetic transcriptions and therefore differ in the number of phones. To effectively handle such variations, we have used dynamic programming with back-tracking technique to align the two phone label sequences (and their U-V and V-U transitions). First, we map each phone in the phone label sequence to either V or U. Next, we note down the time instants, where the transitions from V to U and from U to V occur. We then use dynamic programming to obtain the optimal path to align the two sequences. Next, we traverse along this optimal path and construct a mapping table, which maps the transition instants in the source waveform to those in the target waveform. Using this mapping table, we can now pick the 'V' segments from the source signal and 'U' segments from the target signal and stitch them together to finally construct the mixed signal. The mixed signal is smoothed at the points of stitching to avoid discontinuities in the signal. This entire procedure is illustrated in Fig. 1.

Figures 2 (a) and (b) show the spectrograms of one of the sentences spoken by a male (source) and a female (target) speaker, respectively. Figure 2 (c) shows the spectrogram of the mixed signal, where the voiced portion is taken from the source signal and the unvoiced, from the target signal.

\subsection{Listening tests by human subjects}
Including 10 original utterances, a total of 25 stimuli are presented to each subject in a random order. After each utterance is played, the subject is asked to decide on his/her perception as 'I hear only one speaker' or 'I hear two speakers'. A total of 50 subjects (25 male + 25 female), in the age group of 18 to 60 years, participated as listeners. The results of the listening tests are presented and discussed in Section 4.

\begin{figure}[htb]
\includegraphics[width=0.45\textwidth,height=0.56\textheight]{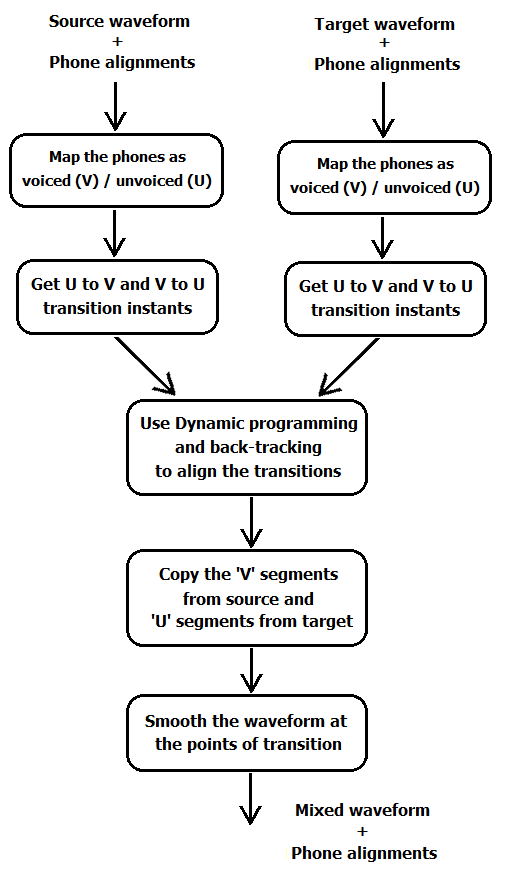}
\centering
\caption{Steps involved in the creation of mixed speech (M$<$FU and F$<$MU), where all the unvoiced segments of the source utterance are replaced by those of the target utterance.}
	\label{fig01}
\end{figure}

\begin{figure}[htb]
\includegraphics[width=0.49\textwidth,height=0.43\textheight]{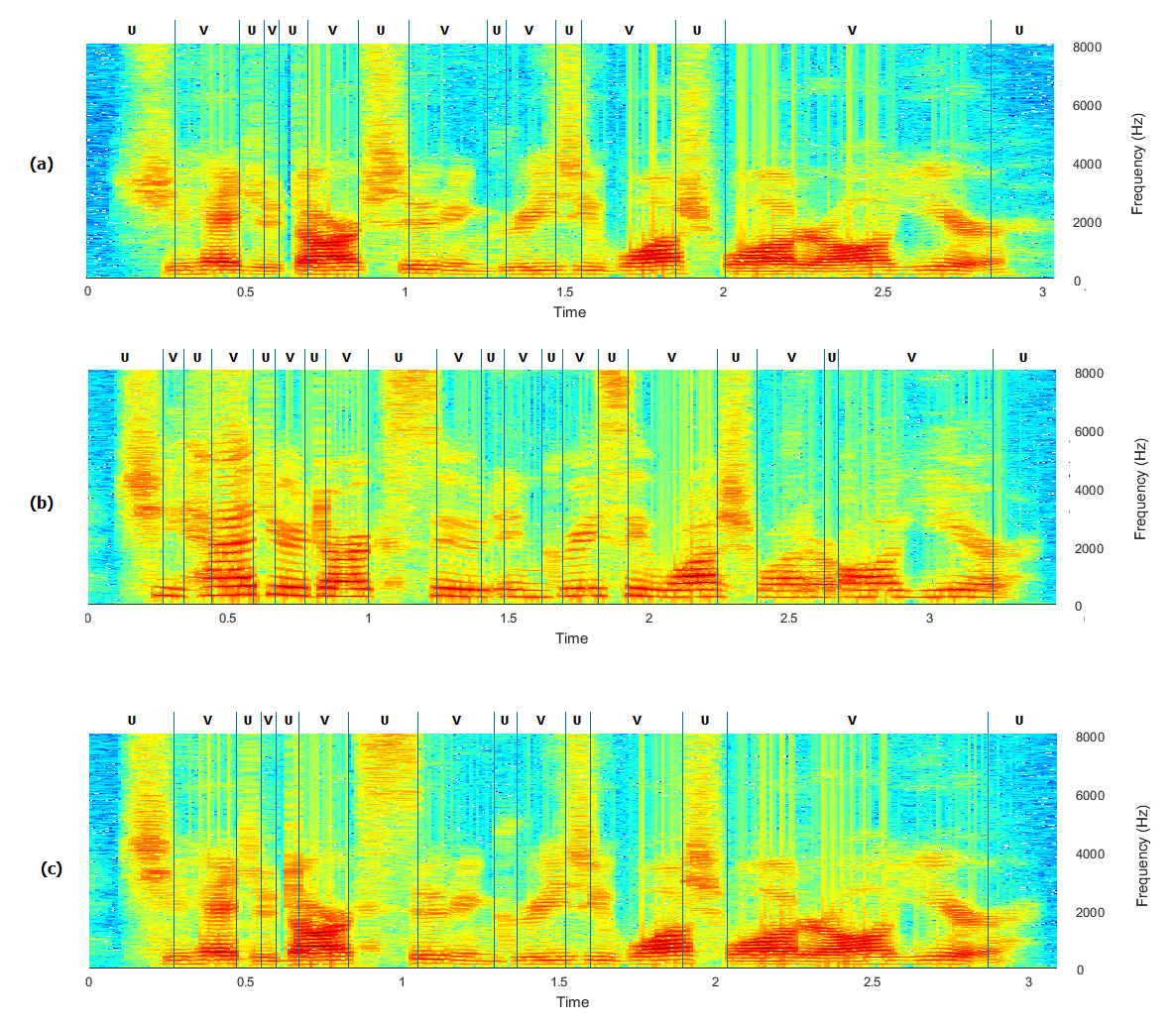}
\centering
\caption{Spectrograms of (a) a source male utterance, (b) a target female utterance of the same sentence, and (c) mixed utterance, with all the unvoiced segments of the male utterance replaced by those of the female, (M$<$FU).}
	\label{fig01}
\end{figure}

\section{Phoneme Recognition Experiments}
All the phoneme recognition system models described below have been trained using the Kaldi toolkit~\cite{kaldi}.

\subsection{Training of Acoustic models}
In the TIMIT training set, there are a total of 3260 and 1360 utterances for male and female speakers, respectively. Using this data, two sets of models have been trained for male and female data, separately. Each model-set comprises the following four models: (i) monophone (Mono), (ii) triphone (Tri), (iii) triphone with LDA+MLLT (LMT) and (iv) triphone with LDA+MLLT+SAT (LMST). The models have been trained in a sequence using force-alignments from the previous stage. Thus, totally eight distinct ASR models have been trained for the objective experiments.

We have used the standard procedure for training the acoustic models as used in the s5 TIMIT Kaldi recipe \cite{kaldi}. First, we extract 13-dimensional mel-frequency cepstral coefficients (MFCC) features with delta and delta-delta MFCC to construct a 39-dimensional feature vector and train a Mono HMM for each of the 48 context-independent phonemes. All HMMs are three state left-to-right models with a total of 1000 Gaussian densities shared among all the states. The second model (Tri) is a set of triphone, context-dependent HMM models trained on the 39-dimensional features with a total of 2500 states and 15000 Gaussians. The third (LMT) is a maximum likelihood, linear transformation based triphone model, trained on 40-dimensional features obtained after splicing and linear transformation (using linear-discriminant analysis). The fourth model (LMST) is a triphone HMM, which is a combination of LDA, MLLT and speaker-adaptive models trained using maximum likelihood linear regression. In all our experiments, we have used tri-gram phone-level language model estimated from the training corpus.

\subsection{Recognition tests on the learnt acoustic models}
We have neglected those sentences which were spoken by only one gender and generated 714 utterances for male and 392, for female. For each of these utterances, a mixed utterance (M$<$FU or F$<$MU; M$<$FvU or F$<$MvU) has been created as described in Sec. 2.1. In addition, to test the relative influence of stops and fricatives, a mixed utterance (M$<$FSSt) is prepared, where only the silence and stop segments of a male speaker's utterance are replaced by the corresponding segments from the utterance of a female speaker and vice-versa (F$<$MSSt). All the utterances, i.e., the original and the mixed, have been tested by automatic speech recognition (ASR) models trained separately on male or female original utterances only. Several experiments are conducted: (i)	Male model is tested on male utterances, (ii)	Male model is tested on female utterances, (iii)	Male model is tested on mixed utterances with male voiced and female unvoiced or stop segments, (iv) Male model is tested on mixed utterances with female voiced and male unvoiced or stop segments; (v) Female model was tested on male utterances; (vi) Female model was tested on female utterances; (vii) Female model is tested on mixed utterances, with male voiced and female unvoiced or stop segments; (viii) Female model is tested on mixed utterances with female voiced and male unvoiced or stop segments.

\section{Results and Discussion}
\subsection{Results of perception experiments}
For the original utterance, the expected correct response is 'I hear only one speaker' and for all the mixed utterances, the expected correct response is 'I hear two speakers'. 
Table 1 lists the mean accuracy of the correct response for all the 25 test utterances by the 50 listeners. The mean accuracy of recognizing the original male utterances by all the listeners is 98.4\%. The corresponding value for the original female utterances is 99.2\%. When all the unvoiced segments (U) and two consecutive voiced segments (v) of the male utterances are replaced by the corresponding segments from female utterances (M$<$FvU), the listeners have correctly identified the presence of two speakers 99.3\% of the time. For a similar mixing of the female utterances (F$<$MvU), listeners identified them correctly 100\% of the time. On the other hand, when only the unvoiced segments of female utterances are replaced (F$<$MU), none of the listeners could identify even one of the mixed utterances as having parts from two different speakers. Thus, the accuracy of correct  respose  of two speakers is 0\%. The corresponding value, when all the unvoiced segments of male utterances are replaced by those of females (M$<$FU), is only 0.4\%. These results clearly indicate that human listeners are unable to detect a change of speaker, when all the unvoiced segments of an utterance (which, usually are nearly 50\% in duration) are replaced by those from the utterance spoken by a speaker of opposite gender.

		\begin{table}
        \centering
		\vspace{0.1cm}
		\caption{Results of perceptual tests conducted on 50 subjects, by presenting the original and modified utterances. M: utterances spoken by males; F: utterances spoken by females; U: all unvoiced segments of an utterance; v: two consecutive voiced segments in the same utterance; vU: Two consecutive voiced segments, in addition to all unvoiced segments; M$<$FU (F$<$MU): U of M (F) replaced by U of female (male); M$<$FvU (F$<$MvU): vU of M (F) replaced by vU of F (M).}
		\label{tab01}
		\resizebox{0.32\textwidth}{0.11\textheight}{ 
		\begin{tabular}{|c|c|} \hline 
			\textbf{Test data} & \textbf{Accuracy} \\ \hline 
			M &  98.4 \\ [5pt]
            \hline 
			F &  99.2 \\ [5pt]
            \hline 
			M$<$FvU &  99.3 \\ [5pt]
            \hline 
            F$<$MvU & 100.0 \\ [5pt]
            \hline 
			M$<$FU  &   0.4 \\ [5pt]
            \hline 
			F$<$MU &   0.0 \\ [5pt]
            \hline 
		\end{tabular}
		}
	\end{table}
    
\subsection{Phone recognition results}
Table 2 summarizes the recognition results obtained by the 8 different ASR models (4 trained on male speech and the others, trained on female speech). Test data consists of (a) the original test utterances from male speakers and (b) their modifications by the replacement of segments (SSt, U or vU) by the corresponding segments of female speakers. Table 3 summarizes the recognition results obtained by the 8 different ASR models on the test utterances from female speakers and their different modifications by the replacement of segments (SSt, U or vU) by the corresponding segments of male speakers.
The results obtained on the male and female utterances are discussed separately. 

The phone error rates (PER) are the lowest, when the original male utterances are recognized by models trained only on male data (left half of Table.2). The PER of 25.3\% obtained by the Mono model reduces gradually with the sophistication in the model to triphone, LMT and LMST, and reaches the lowest value of 16.8\%. The first four entries (left half of Table.2) in the second to fourth rows also show that the performance systematically degrades as more and more female data replaces the segments of the male utterance. 

The PER of the best ASR model (LMST) degrades by about 1.4\% by the replacement of stops (M$<$FSSt), whereas by the replacement of all the U segments (both stops and fricatives) (M$<$FU), the accuracy degrades by 1.8\%, a difference of only about 0.4\%. In other words, the influence of gender on stop consonants is much more significant (nearly 4 times) than on fricatives. This may appear contrary to the postulate of the invariance of stop consonants \cite{inv08, inv09}. However, it may be argued that the ASR results, which are based on MFCCs, cannot be generalized. Replacement of even a couple of sonorant segments brings down the relative accuracy by 4.4\%, whereas when all the U segments are replaced, the drop in accuracy is only 1.8\%. 

Referring to the right half of Table 2, as  expected, the recognition accuracies for the four ASR models trained on female speech are significantly lower than the values obtained by the models trained on male speakers' data. Referring to rows 2-4, when the same male (original or modified) utterances are recognized by ASR models trained on female speech, the general trend is the opposite. In other words, the PER reduces with increased proportion of female speech in the utterances. However, there are exceptions, and in the case of Tri, LMT and LMST models, in some of the cases, the PER increases when more segments are replaced by segments from female speech. However, the error rates are far higher than the corresponding values for the male ASR models. The results are anomalous, since for the case of M$<$FU, the result is better than the case of M$<$FSSt.

\begin{table}[!htbp]
  \caption {Phone error rates for the phoneme recognition system for matched, mismatched and mixed test data created from male speech. The different models used for the ASR experiments are: Mono: monophone; Tri: tripohone; LMT:LDA+MLLT triphone; LMST: LDA+MLLT+SAT triphone. SSt: Silent and stop consonant segments of any utterance; M$<$FSSt: SSt of M replaced by SSt of F. The other notations are the same as those used in Table 1.}

\centering 
\resizebox{0.48\textwidth}{0.074\textheight}{ \begin{tabular}{| c | c c c c | c c c c | }
 \hline
  \multirow{3}{*}{\textbf{Test data}} & \multicolumn{4}{c|}{\textbf{HMM trained on male speech}} & \multicolumn{4}{c|}{\textbf{HMM trained on female speech}} \\[5pt]
  \cline{2-9}
  & Mono & Tri & LMT & LMST & Mono & Tri & LMT  & LMST \\[5pt]
  \hline
  \hline
  M & 25.3 & 19.9 & 18.2 & 16.8 & 39.2 & 37.4 & 36.1 & 26.1 \\[5pt]
  \hline
  M$<$FSSt & 26.7 & 21.8 & 19.6 & 18.2 & 38.7 & 38.0 & 36.5 & 27.6 \\[5pt]
  \hline
  M$<$FU & 27.7 & 22.7 & 20.4 & 18.6 & 38.1 & 37.2 & 35.5 & 26.9 \\[5pt]
\hline
  M$<$FvU& 29.8 & 26.3 & 25.9 & 23.2 & 36.5 & 36.1 & 33.8 & 28.9 \\[5pt]
  \hline
\end{tabular}%
  }
\end{table}

The results for female utterances (Table 3, right half) shows that the lowest phone error rate is obtained when the original female utterances are recognized by models trained only on female speech. The PER of 26.0\% obtained by the Mono model reduces gradually with the sophistication in the model to triphone, LMT and LMST, and reaches the lowest value of 18.6\%. The degradation in accuracy due to the replacement of stops by those of male speakers is about 2.1\%, whereas the contribution to degradation by fricatives is only about 0.8\%. This once again indicates the significant influence of gender on the stop consonants.

\begin{table}[!htbp]
  \caption {Phone error rates for the phoneme recognition system for matched, mismatched and mixed test data, created from female speech. F$<$MSSt: SSt of F replaced by SSt of M. The other notations are the same as those used in Table 1.}

\centering 
\resizebox{0.48\textwidth}{0.074\textheight}{ \begin{tabular}{| c | c c c c | c c c c | }
 \hline
  \multirow{3}{*}{\textbf{Test data}} & \multicolumn{4}{c|}{\textbf{HMM trained on male speech}} & \multicolumn{4}{c|}{\textbf{HMM trained on female speech}} \\[5pt]
  \cline{2-9}
  & Mono & Tri & LMT & LMST & Mono & Tri & LMT  & LMST \\[5pt]
  \hline
  \hline
  F & 39.2 & 37.0 & 39.2 & 25.1 & 26.0 & 22.2 & 20.7 & 18.6 \\[5pt]
  \hline
  F$<$MSSt& 38.8 & 36.4 & 38.9 & 26.3 & 27.2 & 24.2 & 22.1 & 20.7 \\[5pt]
  \hline
 F$<$MU & 37.1 & 34.7 & 37.5 & 25.7 & 29.1 & 25.4 & 23.9 & 21.5 \\[5pt]
  \hline
  F$<$MvU& 36.4 & 33.8 & 36.0 & 27.9 & 31.5 & 29.5 & 27.5 & 26.3 \\[5pt]
\hline
\end{tabular}%
  }
\end{table}

\section{Conclusion}
In order to further our understanding of the influence of speaker's gender on unvoiced sounds, we have conducted both subjective and objective experiments. When the acoustic signal of all the unvoiced segments is replaced by the corresponding signal of a speaker of opposite gender, listeners are unable to detect the speaker change. However, objective experiments based on the PER of a phoneme recognition system show that the speaker's gender does influence the unvoiced sounds. Based on these apparently contradictory findings, we infer the following: (i) listeners may be using speaker-independent acoustic cues, other than MFCCs, for identifying unvoiced sounds; (ii) listeners may not use unvoiced sounds for identifying the gender of a speaker; and (iii) listeners may not be making use of any speaker normalization strategy for unvoiced sounds.

The phone error rates on the modified utterances are slightly higher than those when the ASR is trained and tested on the samples of the same gender. The PER of the ASR increases monotonically with the increase in the proportion of segments replaced by those of the opposite sex. The degradation due to the replacement of stop segments is more significant than that of fricatives. The results also indicate that the speaker recognition systems must give higher weightage to modeling the speech production mechanism, whereas the speech recognition systems ought to give more weightage to modeling the speech perception mechanism.

\bibliographystyle{IEEEtran}

\bibliography{mybib}

\end{document}